\documentclass[10pt,a4paper]{llncs}

\usepackage{makeidx}  

\usepackage{graphics}
\usepackage{epsfig}
\usepackage{graphicx}
\usepackage{xcolor}
\usepackage{verbatim}
\usepackage{amsfonts}
\usepackage{times}
\usepackage{amstext}
\usepackage{amsmath}
\usepackage{algorithm}

\catcode`\.=\active\def.{\char'56\allowbreak}\catcode`\.=12
\catcode`\/=\active\def/{\char'57\allowbreak}\catcode`\/=12
\catcode`\:=\active\def:{\char'72\allowbreak}\catcode`\:=12  
\catcode`\@=\active\def@{\char'100\allowbreak}\catcode`\@=12 
\catcode`\-=\active\def-{\char'55\allowbreak}\catcode`\-=12  
\def\URL{\bgroup\catcode`\.=\active\catcode`\/=\active\catcode`\:=\active\catcode`\@=\active\catcode`\-=\active\def~{\char126}\tt\URLaux}
\def\URLaux#1{#1\egroup}

\title{Fractionally Predictive Spiking Neurons}
\titlerunning{Fractionally Predictive Spiking Neurons}

\author{Sander Bohte \& Jaldert Rombouts}
\authorrunning{S.M. Bohte \& J.O. Rombouts}   
\tocauthor{Sander Bohte}

\institute{CWI, Center for Mathematics and Computer Science\\
Science Park 123, NL-1098 XG Amsterdam, The Netherlands\\
\{S.M.Bohte,J.O.Rombouts\}@cwi.nl\\
}

\begin{document}

\maketitle

\begin{abstract}
Recent experimental work has suggested that the neural firing rate can
be interpreted as a fractional derivative, at least when signal
variation induces neural adaptation. Here, we show that the actual
neural spike-train itself can be considered as the fractional
derivative, provided that the neural signal is approximated by a sum
of power-law kernels. A simple standard thresholding spiking neuron suffices to carry out such an approximation, given a suitable refractory response. 
Empirically, we find that the online
approximation of signals with a sum of power-law kernels is beneficial
for encoding signals with slowly varying components, like long-memory
self-similar signals. For such signals, the online power-law kernel approximation typically required
less than half the number of spikes for similar SNR as compared to sums of
similar but exponentially decaying kernels. 
As power-law kernels can be accurately approximated using sums or
cascades of weighted exponentials, we demonstrate that the
corresponding decoding of spike-trains by a receiving neuron allows for natural and transparent temporal signal filtering by tuning the weights of the
decoding kernel.
\end{abstract}

\section{Introduction}

A key issue in computational neuroscience is the interpretation of neural signaling, as expressed by a neuron's sequence of action potentials. 
An emerging notion is that neurons may in fact encode information at multiple timescales simultaneously \cite{fairhall2001multiple,wark2009timescales,panzeri2010sensory,lundstrom2010multiple}: the precise timing of spikes may be conveying high-frequency information, and slower measures, like the rate of spiking, may be relating low-frequency information. Such multi-timescale encoding comes naturally, at least for sensory neurons, as the statistics of the outside world often exhibit self-similar multi-timescale features \cite{vanHateren1997processing} and the magnitude of natural signals can extend over several orders. 
Since neurons are limited in the rate and resolution with which they can emit spikes, the mapping of large dynamic-range signals into spike-trains is an integral part of attempts at understanding neural coding. 

Experiments have extensively demonstrated that neurons adapt their response when facing persistent changes in signal magnitude. Typically, adaptation changes the relation between the magnitude of the signal and the neuron's discharge rate. Since adaptation thus naturally relates to neural coding, it has been extensively scrutinized
\cite{brenner2000adaptive,wark2007sensory,famulare2009feature}. Importantly, adaptation is found to additionally exhibit features like dynamic gain control, when the standard deviation but not the mean of the signal changes \cite{fairhall2001multiple}, and long-range time-dependent changes in the spike-rate response are found in response to large magnitude signal steps, with the changes following a power-law decay (e.g. \cite{drew2006models}). 

Tying the notions of self-similar multi-scale natural signals and adaptive neural coding together, it has recently been suggested that neuronal adaptation allows neuronal spiking to communicate a {\em fractional derivative} of the actual computed signal \cite{lundstrom2008fractional,lundstrom2010multiple}. Fractional derivatives are a generalization of standard `integer' derivatives (`first order', `second order'),  to real valued derivatives (e.g. `0.5th order'). A key feature of such derivatives is that they are non-local, and rather convey information over essentially a large part of the signal spectrum \cite{lundstrom2008fractional}. 

Here, we show how neural spikes can encode temporal signals when the spike-train {\em itself} is taken as the fractional derivative of the signal. We show that this is the case for a signal approximated by a sum of shifted power-law kernels starting at respective times $t_i$ and decaying proportional to $1/(t-t_i)^{\beta}$. 
Then, the fractional derivative of this approximated signal corresponds to a sum of spikes at times $t_i$, provided that the order of fractional differentiation $\alpha$ is equal to  $1-\beta$: a spike-train {\em{is}} the $\alpha = 0.2$ fractional derivative of a signal approximated by a sum of power-law kernels with exponent $\beta = 0.8$. Such signal encoding with power-law kernels can be carried out for example with simple standard thresholding spiking neurons with a refractory reset following a power-law. 


As fractional derivatives contain information over many time-ranges, they are naturally suited for predicting signals. This links to notions of predictive coding, where neurons communicate deviations from expected signals rather than the signal itself. Predictive coding has been suggested as a key feature of neuronal processing in e.g. the retina \cite{hosoya2005dynamic}. For self-similar scale-free signals, future signals may be influenced by past signals over very extended time-ranges: so-called long-memory. For example, fractional Brownian motion (fBm) can exhibit long-memory, depending on their Hurst-parameter $H$. For $H>0.5$ fBM models which exhibit long-range dependence (long-memory) where the autocorrelation-function follows a power-law decay \cite{wornell1999signal}. The long-memory nature of signals approximated with sums of power-law kernels naturally extends this signal approximation into the future along the autocorrelation of the signal, at least for self-similar $1/f^{\gamma}$ like signals. The key ``predictive'' assumption we make is that a neuron's spike-train up to time $t$ contains all the information that the past signal contributes to the future signal $t'>t$.  




The correspondence between a spike-train as a fractional derivative and a signal approximated as a sum of power-law kernels is only exact when spike-trains are taken as a sum of Dirac-$\delta$ functions and the power-law kernels as $1/t^{\beta}$. As both responses are singular, neurons would only be able to approximate this. We show empirically how sums of (approximated) $1/t^{\beta}$ power-law kernels can accurately approximate long-memory fBm signals via simple difference thresholding, in an online greedy fashion. Thus encodings signals, we show that the power-law kernels approximate synthesized signals with about half the number of spikes to obtain the same Signal-to-Noise-Ratio, when compared to the same encoding method using similar but exponentially decaying kernels. 

We further demonstrate the approximation of sine wave modulated white-noise signals with sums of power-law kernels. The resulting spike-trains, expressed as  ``instantaneous spike-rate'', exhibit the phase-presession as in \cite{lundstrom2010multiple}, with suppression of activity on the ``back'' of the sine-wave modulation, and stronger suppression for lower values of the power-law exponent (corresponding to a higher order for {\em our} fractional derivative). 
We find the effect is stronger when encoding the actual sine wave envelope, mimicking the difference between thalamic and cortical neurons reported in \cite{lundstrom2010multiple}. This may suggest that these cortical neurons are more concerned with encoding the sine wave envelope.



The power-law approximation also allows for the transparent and straightforward implementation of temporal signal filtering by a post-synaptic, receiving neuron. Since neural {\em de}coding by a receiving neuron corresponds to adding a power-law kernel for each received spike, modifying this receiving power-law kernel then corresponds to a temporal filtering operation, effectively exploiting the wide-spectrum nature of power-law kernels. This is particularly relevant, since, as has been amply noted \cite{drew2006models,fusi2005cascade}, power-law dynamics can be closely approximated by a weighted sum or cascade of exponential kernels. Temporal filtering would then correspond to simply tuning the weights for this sum or cascade. We illustrate this notion with an encoding/decoding example for both a high-pass and low-pass filter.

\begin{figure}[t]
\center
\includegraphics[width=110mm,height=40mm]{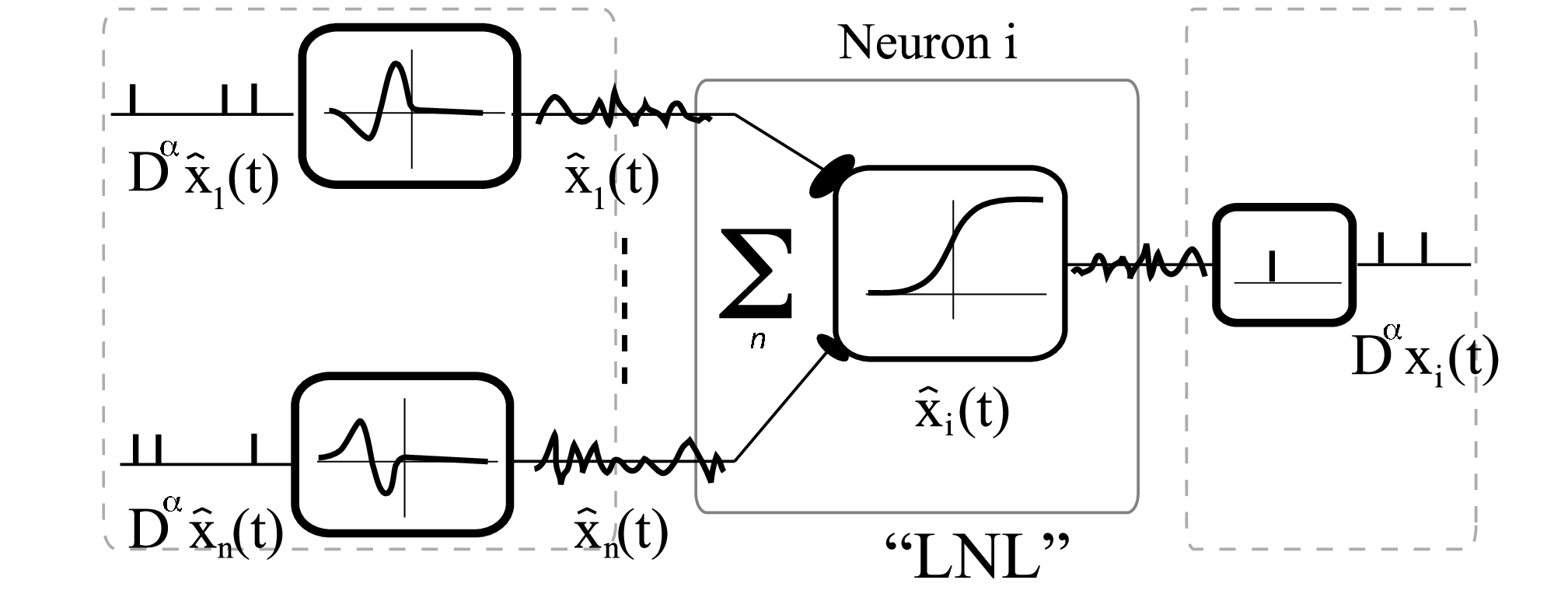}
\caption{Linear-Non-Linear filter, with spike-decoding front-end and spike-encoding back-end. 
}
\label{fig:LNL}
\end{figure}

\vspace{-0.00025cm}
\section{Power-law Signal Encoding}
\vspace{-0.0002cm}
Neural processing can often be reduced to a Linear-Non-Linear (LNL) filtering operation on incoming signals \cite{bishop1995neural} (figure \ref{fig:LNL}), where inputs are linearly weighted and then passed through a non-linearity to yield the neural activation. As this computation yields analog activations, and neurons communicate through spikes, the additional problem faced by spiking neurons is to decode the incoming signal and then encode the computed LNL filter again into a spike-train. The standard spiking neuron model is that of Linear-Nonlinear-Poisson spiking, where spikes have a stochastic relationship to the computed activation \cite{chichilnisky2001simple}. Here, we interpret the spike encoding and decoding in the light of processing and communicating signals with fractional derivatives \cite{lundstrom2008fractional}.

At least for signals with mainly (relatively) high-frequency components, it has been well established that a neural signal can be decoded with high fidelity by associating a fixed kernel with each spike, and summing these kernels \cite{rieke1999spikes}; keeping track of doublets and triplet spikes allows for even greater fidelity. This approach however only worked for signals with a frequency response lacking low frequencies \cite{rieke1999spikes}. Low-frequency changes lead to ``adaptation'', where the kernel is adapted to fit the signal again \cite{fairhall2001efficiency}. For long-range predictive coding, the absence of low frequencies leaves  little to predict, as the effective correlation time of the signals is then typically very short as well \cite{rieke1999spikes}. 

Using the notion of predictive coding in the context of (possible) long-range dependencies, we define the goal of signal encoding as follows: let a signal $x_j(t)$ be the result of the continuous-time computation in neuron $j$ up to time $t$, and let neuron $j$ have emitted spikes $t_j$ up to time $t$. These spikes should be emitted such that the signal $x_j(t')$ for $t'<t$ is decoded up to some signal-to-noise ratio, {\em and} these spikes should be predictive for $x_j(t')$ for $t'>t$ in the sense that no additional spikes are needed at times $t'>t$ to convey the predictive information up to time $t$. 


Taking kernels as a signal filter of fixed width, as in the general approach in \cite{rieke1999spikes} has the important drawback that the signal reconstruction incurs a delay for the duration of the filter: its detection cannot be communicated until the filter is actually matched to the signal. This is inherent to any backward-looking filter-maching solution. Alternatively, a predictive coding approach could rely on only on a very short backward looking filter, minimizing the delay in the system, and continuously computing a forward predictive signal. At any time in the future then, only deviations of the actual signal from this expectation are communicated. 




\subsection{Spike-trains as fractional derivative}
As recent work has highlighted the possibility that neurons encode fractional derivatives, it is noteworthy that the non-local nature of fractional calculus offers a natural framework for predictive coding. In particular, as we will show, when we assume that the predictive information about the future signal is fully contained in the current set of spikes, a signal approximated as a sum of power-law kernels corresponds to a fractional derivative in the form of a sum of Dirac-$\delta$ functions, which the neuron can obviously communicate through timed spikes.

The fractional derivative $r(t)$ of a signal $x(t)$ is denoted as $D^{\alpha} x(t) $, and intuitively expresses:
\[
r(t) = \frac{d^{\alpha}}{d t^{\alpha}} x(t),
\]
where $\alpha$ is the fractional order, e.g. $0.5$. This is most conveniently computed through the Fourier transformation in the frequency domain, as a simple multiplication:
\[
R(\omega) = H(\omega) X(\omega), 
\]
where the Fourier-transformed fractional derivative operator $H(\omega)$ is by definition $(i\omega)^{\alpha}$ \cite{lundstrom2008fractional}, and $X(\omega)$ and $R(\omega)$ are the Fourier transforms of $x(t)$ and $r(t)$ respectively. 

We assume that neurons carry out predictive coding by emitting spikes such that all predictive information is contained in the current spikes, and no more spikes will be fired if the signal follows this prediction. Approximating spikes by Dirac-$\delta$ functions, we take the spike-train up to some time $t_0$ to be the fractional derivative of the past signal {\em and} be fully predictive for the expected influence the past signal has on the future signal:
\[
r(t) = \sum_{t_i < t_0} \delta(t-t_i)
\]
The task is to find a signal $\hat{x}(t)$ that corresponds to an approximation of the actual signal $x(t)$ up to $t_0$, and where the predicted signal contribution $x(t)$ for $t>t_0$ due to $x(t<t_0)$ does not require additional future spikes. We note that a sum of power-law decaying kernels with power-law $t^{-\beta}$ for $\beta = 1-\alpha$ corresponds to such a fractional derivative: the Fourier-transform for a power-law decaying kernel of form $t^{-\beta}$ is proportional to $(i\omega)^{\beta-1}$, hence for a signal that just experienced a single step from 0 to 1 at time $t$ we get:
\[
R(\omega) =  (i\omega)^{\alpha} (i\omega)^{\beta-1}, 
\]
and setting $\beta =  1-\alpha$ yields a constant in Fourier-space, which of course is the Fourier-transform of $\delta(t)$. It is easy to check that shifted power-law decaying kernels, e.g.  $(t-t_a)^{-\beta}$ correspond to a shifted fractional derivative $\delta(t-t_a)$, and the fractional derivative of a sum of shifted power-law decaying kernels corresponds to a sum of shifted delta-functions. Note that for decaying power-laws, we need $\beta >0$, and for fractional derivatives we require $\alpha >0$.

Thus, with the reverse reasoning, a signal approximated as the sum of power-law decaying kernels corresponds to a spike-train with spikes positioned at the start of the kernel, and, beyond a current time $t$, this sum of decaying kernels is is interpreted as a prediction of the extent to which the future signal can be predicted by the past signal.

\begin{figure}[t]
\center
\includegraphics[width=120mm]{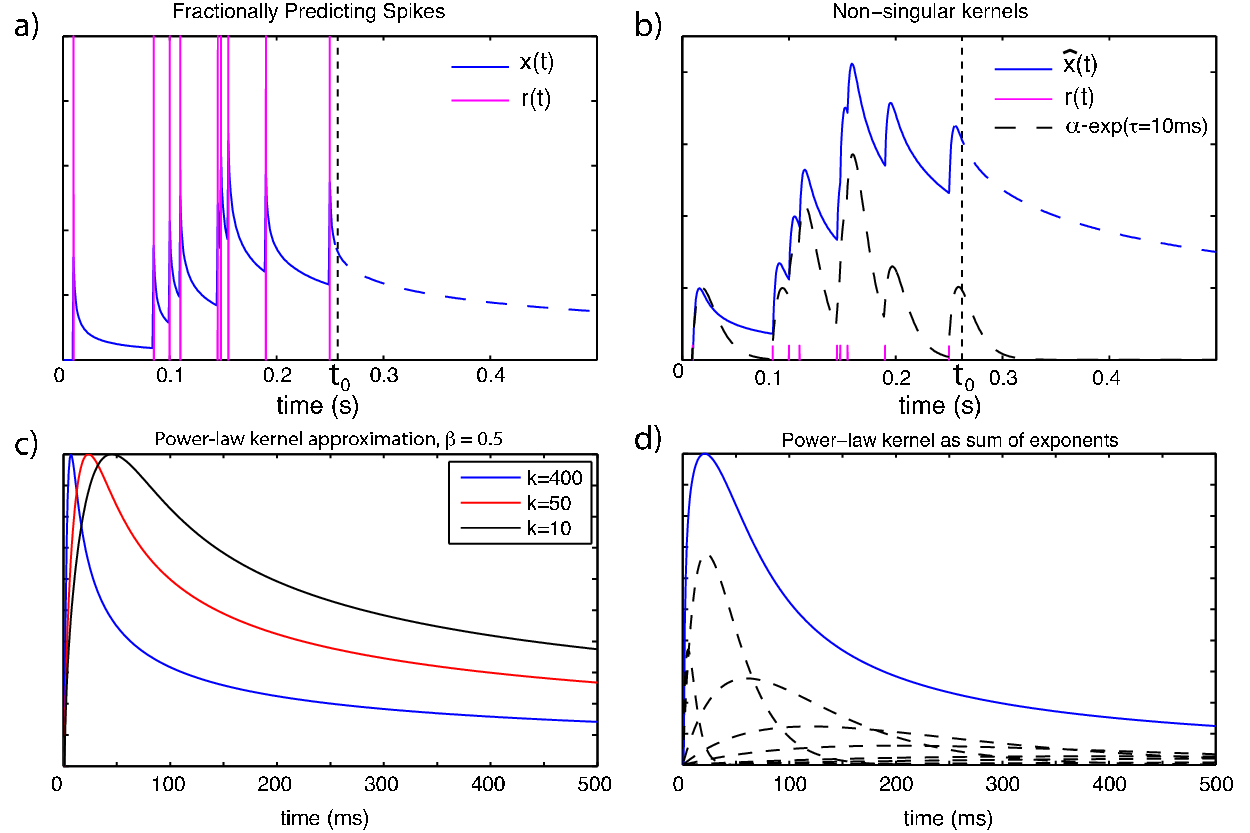}
\caption{a) Signal $x(t)$ and corresponding fractional derivative $r(t)$: $1/t^{\beta}$ power-laws and delta-functions; b) power-law approximation, timed to spikes; compared to sum of $\alpha$-functions (black dashed line). c) Approximated $1/t^{\beta}$ power-law kernel for different values of $k$ from eq. (2). d) The approximated $1/t^{\beta}$ power-law kernel (blue line) can be decomposed as a weighted sum of $\alpha$-functions with various decay time-constants (dashed lines).
\vspace{-0.00025cm}
}
\label{fig:approxkern}
\end{figure}

Obviously, both the Dirac-$\delta$ function and the $1/t^{\beta}$ kernels are singular (figure \ref{fig:approxkern}a) and can only be approximated. For real applications, only some part of the $1/t^{\beta}$ curve can be considered, effectively leaving the magnitude of the kernel and the high frequency component (the extend to which the initial $1/t^{\beta}$ peak is approximated) as free parameters. Figure \ref{fig:approxkern}b illustrates the signal approximated by a random spikes train; as compared to a sum of exponentially decaying $\alpha$-kernels, the long-memory effects of power-law decay kernels is evident. 

\subsection{Practical encoding}
To explore the efficacy of the power-law kernel approach to signal encoding/decoding, we take a standard thresholding online approximation approach, where neurons communicate only deviations between the current computed signal $x(t)$ and the emitted approximated signal $\hat{x}(t)$ exceeding some threshold $\theta$. 
The  emitted signal $\hat{x}(t)$ is constructed as the (delayed) sum of filter kernels $\kappa$ each starting at the time of the emitted spike:
\[
\hat{x}(t) = \sum_{t_j < t} \kappa(t-(t_j+\Delta)),
\]
the delay $\Delta$ corresponds to the time-window over which the neuron considers the difference between computed and emitted signal. In a spiking neuron, such computation would be implemented simply by for instance a refractory current following a power-law. Allowing for both positive and negative spikes (corresponding to tightly coupled neurons with reversed threshold polarity \cite{rieke1999spikes}), this would expand to:
\[
\hat{x}(t) = \sum_{t_j^+ < t} \kappa(t-(t_j^+ +\Delta))-\sum_{t_j^- < t} \kappa(t-(t_j^- +\Delta)).
\]
Considering just the fixed time-window thresholding approach, a spike is emitted each time the difference between the computed signal $x(t)$ and the emitted signal $\hat{x}(t)$ plus (or minus) the kernel $\kappa(t)$ summed over some time-window exceeds the threshold $\theta$: 
\begin{align}
r(t_0) & = \delta(t_0) & \quad \text{if} \sum_{\tau=t_0-\Delta}^{t_0} |x(\tau)-\hat{x}(\tau)| - |x(\tau)-(\hat{x}(\tau)+\kappa(\tau))|) > \theta, \nonumber \\
	 & =  -\delta(t_0) & \quad \text{if} \sum_{\tau=t_0-\Delta}^{t_0} |x(\tau)-\hat{x}(\tau)| - |x(\tau)-(\hat{x}(\tau)-\kappa(\tau))|) > \theta,
\end{align} 
the signal approximation improvement is computed here as the absolute value of the difference between the current signal noise and the signal noise when a kernel is added (or subtracted).



As an approximation of $1/t^{\beta}$ power-law kernels, we let the kernel first quickly rise, and then decay according to the power-law. For a practical implementation, we use a $1/t^{\beta}$ signal multiplied by a modified version of the logistic sigmoid function $\text{logsig}(t) = 1 / (1 + \exp(-t))$: $v(t,k) = 2 \,\text{logsig}(k t)-1$, such that the kernel becomes:
\begin{equation}
\kappa(t) = \lambda v(t,k) 1/t^{\beta},
\label{eq:plaw}
\end{equation}
where $\kappa(t)$ is zero for $t'<t$, and parameter $k$ determines the angle of the initial increasing part of the kernel. The resulting kernel is further scaled by a factor $\lambda$ to achieve a certain signal approximation precision (kernels for power-law exponential $\beta = 0.5$ and several values of $k$ are shown in figure \ref{fig:approxkern}c). As an aside, the resulting (normalized) power-law kernel can very accurately be approximated over multiple orders of magnitude by a sum of just 11 $\alpha$-function exponentials (figure \ref{fig:approxkern}d).
 
Next, we compare the efficiency of signal approximation with power-law predictive kernels as compared to the same approximation using standard fixed kernels. For this, we synthesize self-similar signals with long-range dependencies. We first remark on some properties of self-similar signals with power-law statistics, and on how to synthesize them.

\subsection{Self-similar signals with power-law statistics}
There is extensive literature on the synthesis of statistically self-similar signals with $1/f$-like statistics, at least going back to Kolmogorov \cite{kolmogorov1940kurven} and Mandelbrot \cite{mandelbrot1968fractional}. Self-similar signals exhibit slowly decaying variances, long-range dependencies and a spectral density following a power law. Importantly, for wide-sense self-similar signals, the autocorrelation functions also decays following a power-law.
Although various distinct classes of self-similar signals with $1/f$-like statistics exist \cite{wornell1999signal}, fractional Brownian motion (fBm) is a popular model for many natural signals. Fractional Brownian motion is characterized by its Hurst-paramater $H$, where $H=0.5$ corresponds to regular Brownian motion, and fBM models with $H>0.5$ exhibit long-range (positive) dependence. The spectral density of an fBm signal is proportional to a power-law, $1/f^{\gamma}$, where $\gamma = 2H+1$. 
We used fractional Brownian motion to generate self-similar signals for various $H$ values, using the {\tt wfbm} function from the Matlab wavelet toolbox.

\section{Signal encoding/decoding}
\vspace{-0.00025cm}
\subsection{Encoding long-memory self-similar signals}
\begin{figure}[t]
\center
\includegraphics[width=140mm]{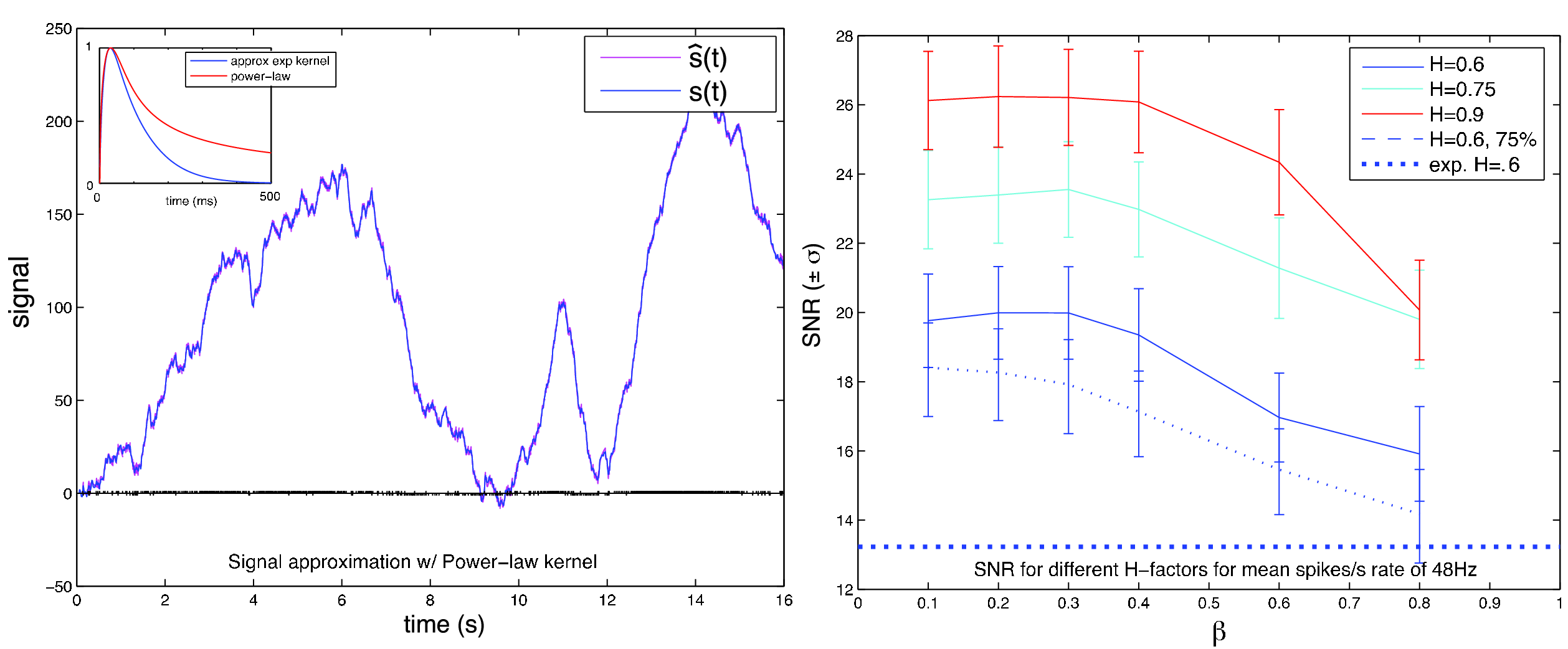}
\caption{Left: example of encoding of fBm signal with power-law kernels. Using an exponentially decaying kernel (inset) required 1398 spikes vs. 618 for the power-law kernel ($k=50$), for the same SNR. Right: SNR for various $\beta$ power-law exponents using a fixed number of spikes (48Hz), with curves for different $H$-parameters, each curve averaged over five 16s signals. The dashed blue curve plots the $H=0.6$ curve, using less spikes (36Hz); the flat bottom dotted line shows the average performance of the non-power-law exponentially decaying kernel, also for $H=0.6$.  
\vspace{-0.00025cm}
}
\label{fig:approxsign}
\end{figure}

We applied the thresholded kernel approximation outlined above to synthesized fBm signals with $H>0.5$, to ensure long-term dependence in the signal. An example of such encoding is given in figure \ref{fig:approxsign}, left panel, using both positive and negative spikes, (inset, red line: the power-law kernel used). When encoding the same signal with kernels without the power-law tail (inset, blue line), the approximation required more than twice as many spikes for the same Signal-to-Noise-Ratio (SNR).

In figure \ref{fig:approxsign}, right panel, we compared the encoding efficacy for signals with different  $H$-parameters, as a function of the power-law exponent, using the same number of spikes for each signal (achieved by changing the $\lambda$ parameter and the threshold $\theta$). 
We find that more slowly varying signals, corresponding to higher $H$-parameters, are better encoded by the power-law kernels,  
More surprisingly, we find 
and signals are consistently best encoded for low $\beta$-values, in the order of $0.1 - 0.3$.
Similar results were obtained for different values of $k$ in equation \eqref{eq:plaw}.

We should remark that without negative spikes, there is no longer a clear performance advantage for power-law kernels (even for large $\beta$): where power-law kernels are beneficial on the rising part of a signal, they lose on downslopes where their slow decay cannot follow the signal. 
    
\subsection{Sine-wave modulated white-noise}   
\begin{figure}[t]
\center
\includegraphics[width=140mm,height=50mm]{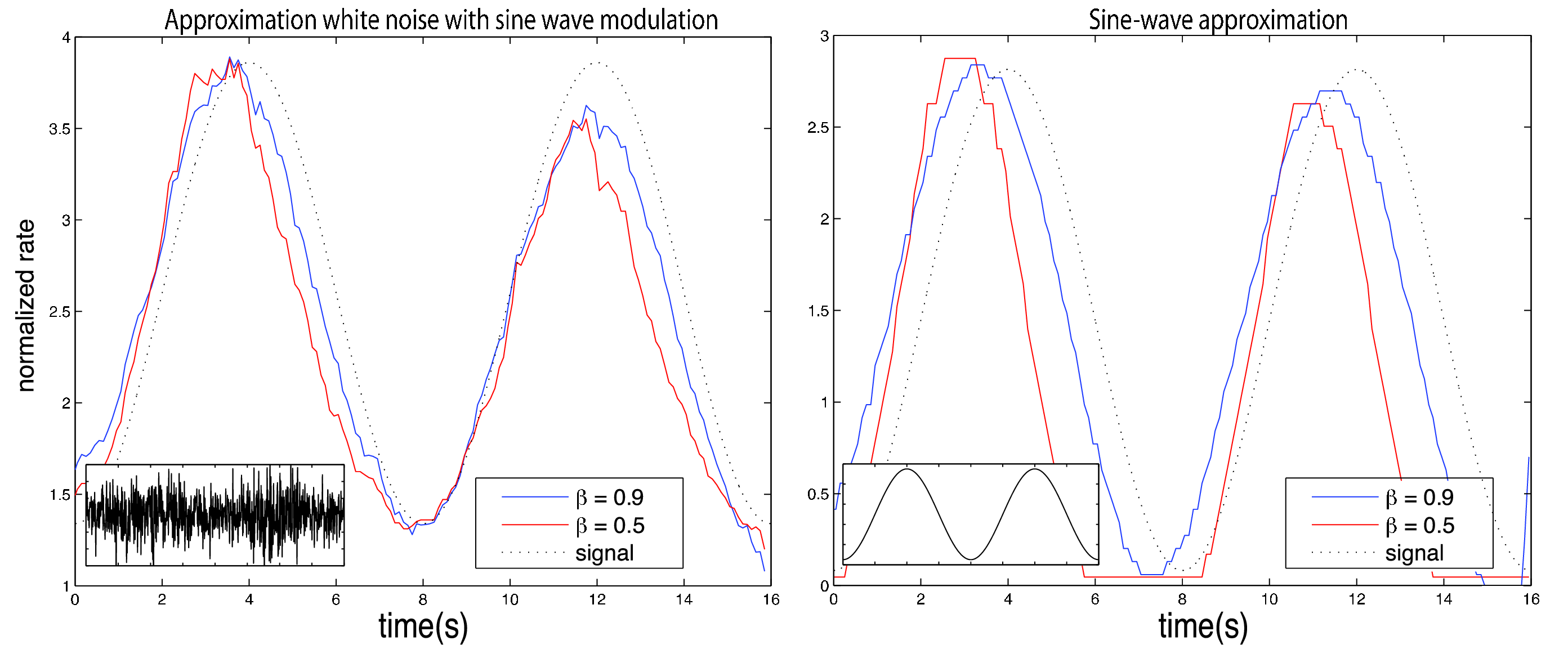}
\caption{Sinewave phase-lead. Left: when encoding sine-wave modulated white noise (inset); right: encoding the sine-wave signal itself (inset). Average firing rate is computed over 100ms, and normalized to match the sine-wave kernel.
\vspace{-0.00035cm}
}
\label{fig:phaselead}
\end{figure}

Fractional derivatives as an interpretation of neuronal firing-rate has been put forward by a series of recent papers \cite{lundstrom2008fractional,lundstrom2009sensitivity,lundstrom2010multiple}, where experimental evidence was presented to suggest such an interpretation. 
A key finding in \cite{lundstrom2010multiple} was that the instantaneous firing rate of neurons along various processing stages of a rat's whisker movement exhibit a  phase-lead relative to the amplitude of the movement modulation. The phase-lead was found to be greater for cortical neurons as compared to thalamic neurons. When the firing rate corresponds to the $\alpha$-order fractional derivative, the phase-lead would correspond to greater fractional order $\alpha$ in the cortical neurons \cite{lundstrom2008fractional} . We used the sum-of-power-laws to approximate both the sine-wave-modulated white noise and the actual sine-wave itself, and found similar results (figure \ref{fig:phaselead}): smaller power-law exponents, in our interpretation also corresponding to larger fractional derivative orders, lead to increasingly fewer spikes at the back of the sine-wave (both in the case where we encode the signal with both positive and negative spikes -- then counting only the positive spikes -- and when the signal is approximated with only positive spikes -- not shown). We find an increased phase-lead when approximating the actual sine-wave kernel as opposed to the white-noise modulation, suggesting that perhaps cortical neurons more closely encode the former as compared to thalamic neurons.

\subsection{Signal Frequency Filtering}
For a receiving neuron $i$ to properly interpret a spike-train $r(t)_j$ from neuron $j$, both neurons would need to keep track of past events over extended periods of time: current spikes have to be added to or subtracted from the future expectation signal that was already communicated through past spikes. The required power-law processes can be implemented in various manners, for instance as a weighted sum or a cascade of exponential processes \cite{drew2006models,lundstrom2008fractional}. 
A natural benefit of implementing power-law kernels as a weighted sum or cascade of exponentials is that a receiving neuron can carry out temporal signal filtering simply by tuning the respective weight parameters for the kernel with which it decodes spikes into a signal approximation. 

\begin{figure}[t]
\center
\includegraphics[width=140mm]{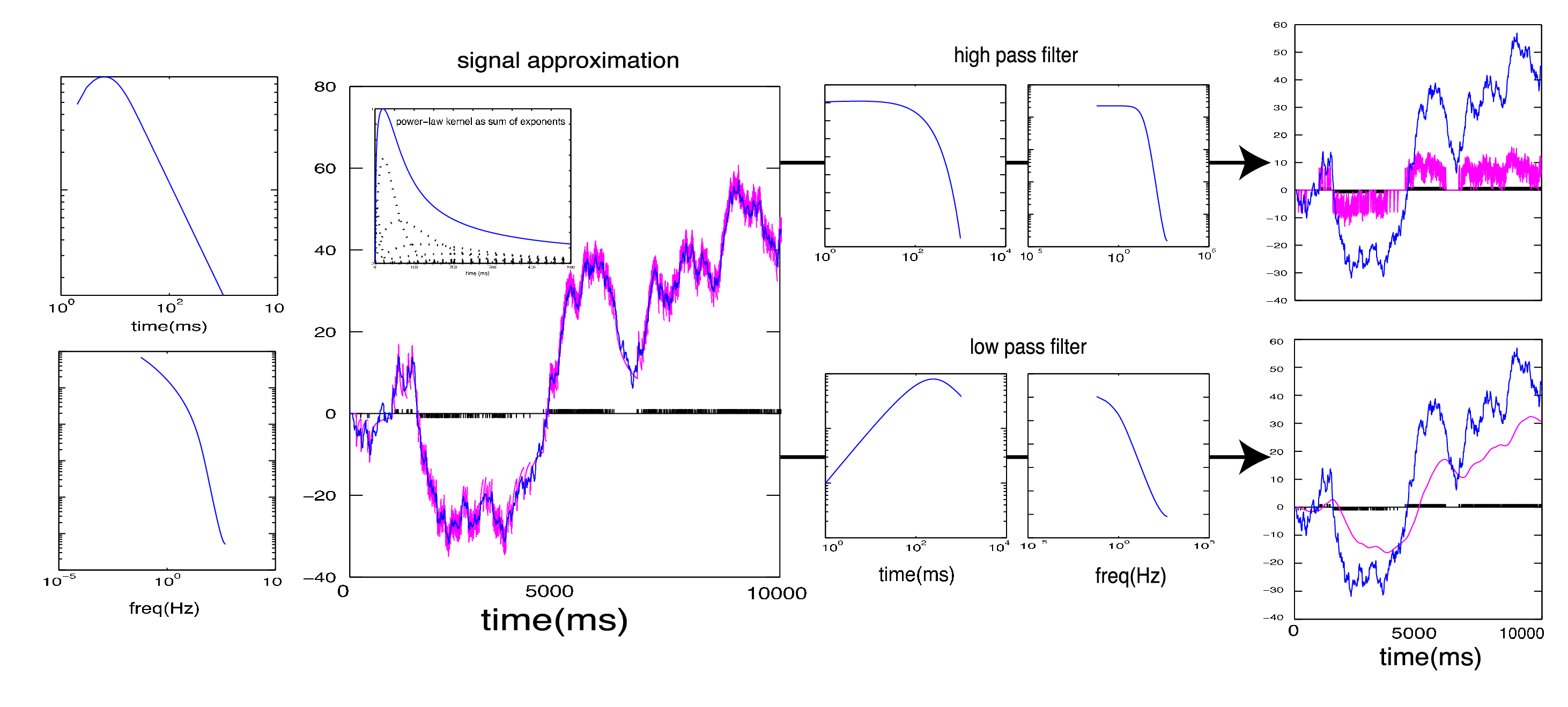}
\caption{Illustration of frequency filtering with modified decoding kernels. The square boxes show the respective kernels in both time and frequency space.  See text for further explanation.
\vspace{-0.0005cm}
}
\label{fig:frequencyfilt2}
\end{figure}

In figure \ref{fig:frequencyfilt2}, we illustrate this with power-law kernels that are transformed into high-pass and low-pass filters. We first approximated our power-law kernel \eqref{eq:plaw} with a sum of 11 exponentials (depicted in the left-center inset). Using this approximation, we encoded the signal (figure \ref{fig:frequencyfilt2}, center). The signal was then reconstructed using the resultant spikes, using the power-law kernel approximation, but with some zeroed out exponentials (respectively the slowly decaying exponentials for the high-pass filter, and the fast-decaying kernels for the low-pass filter). Figure \ref{fig:frequencyfilt2}, most right, shows the resulting filtered signal approximations. Obviously, more elaborate tuning of the decoding kernel with a larger sum of kernels can approximate a vast variety of signal filters. 





%

\section{Discussion}
\vspace{-0.00045cm}

Taking advantage of the relationship between power-laws and fractional derivatives, we outlined the peculiar fact that a sum of Dirac-$\delta$ functions, when taken as a fractional derivative, corresponds to a signal in the form of a sum of power-law kernels. Exploiting the obvious link to spiking neural coding, we showed how a simple thresholding spiking neuron can compute a signal approximation as a sum of power-law kernels; importantly, such a simple thresholding spiking neuron closely fits standard biological spiking neuron models, when the refractory response follows a power-law decay (e.g.  \cite{pozzorini2010}). We demonstrated the usefulness of such an approximation when encoding slowly varying signals, finding that encoding with power-law kernels significantly outperformed similar but exponentially decaying kernels that do not take long-range signal dependencies into account. 

Compared to the work where the firing rate is considered as a fractional derivative, e.g. \cite{lundstrom2008fractional}, the present formulation extends the notion of neural coding with fractional derivatives to individual spikes, and hence finer temporal variations: each spike effectively encodes very local signal variations, while also keeping track of long-range variations. 

The interpretation in \cite{lundstrom2008fractional} of the fractional derivative $r(t)$ as a {\em rate} leads to a 1:1 relation between the fractional derivative order and the power-law decay exponent of adaptation of about $0.2$ \cite{lundstrom2008fractional,xu1996logarithmic,drew2006models}. For such fractional derivative $\alpha$, our derivation implies a power-law exponent for the power law kernels $\beta =  1-\alpha \approx 0.8$, consistent with our sine-wave reconstruction, as well as with recent adapting spiking neuron models \cite{pozzorini2010}. We find that when signals are approximated with non-coupled positive and negative neurons (i.e. one neuron encodes the positive part of the signal, the other the negative), such much faster-decaying power-law kernels encode more efficiently than slower decaying ones. Non-coupled signal encoding obviously fair badly when signals rapidly change polarity; this however seems consistent with human illusory experiences \cite{stocker2009}. 


As noted, the singularity of $1/t^{\beta}$ power-law kernels means that initial part of the kernel can only be approximated. Here, we initially focused our simulation on the use of long-range power-law kernels for encoding slowly varying signals. A more detailed approximation of this initial part of the kernel may be needed to incorporate effects like gain modulation \cite{hong2008intrinsic,famulare2009feature}, and determine up to what extent the power-law kernels already account for this phenomenon. This would also provide a natural link to existing neural models of spike-frequency adaptation, e.g. \cite{jolivet2006integrate}, as they are primarily concerned with modeling the spiking neuron behavior rather than the computational aspects. 

We used a greedy online thresholding process to determine when a neuron would spike to approximate a signal, this in contrast to offline optimization methods that place spikes at optimal times, like Smith \& Lewicki \cite{smith2005efficient}. The key difference of course is that the latter work is concerned with decoding a signal, and in effect attempts to determine the effective neural (temporal) filter. As we aimed to illustrate in the signal filtering example, these notions are not mutually exclusive: a receiving neuron could very well filter the incoming signal with a carefully shaped weighted sum of kernels, and then, when the filter is activated, signal the magnitude of the match through fractional spiking. 

Predictive coding seeks to find a careful balance between encoding known information as well as future, derived expectations \cite{tishby2000information}. It does not seem unreasonable to formulate this balance as a no-going-back problem, where current computations are projected forward in time, and corrected where needed. In terms of spikes, this would correspond to our assumption that, absent new information, no additional spikes need to be fired by a neuron to transmit this forward information. 

The kernels we find are somewhat in contrast to the kernels found by Bialek et. al. \cite{rieke1999spikes}, where the optimal filter exhibited both a negative and a positive part and no long-range ``tail''. Several practical issues may contribute to this difference, not least the relative absence of low frequency variations, as well as the fact that the signal considered is derived from the fly's H1 neurons. These two neurons have only partially overlapping receptive fields, and the separation into positive and negative spikes is thus slightly more intricate. We need to remark though that we see no impediment for the presented signal approximation to be adapted to such situations, or situations where more than two neurons encode fractions of a signal, as in population coding, e.g.  \cite{huys2007fast}.

Finally, we would like to remark that the issue of long-range temporal dependencies such as discussed here seems to be relatively unappreciated. As pointed out in \cite{drew2006models}, long-range power-law dynamics would seem to offer a variety of ``hooks'' for computation through time, like for temporal difference learning and relative temporal computations (and possibly exploiting the many noted correspondences between spatial and temporal statistics \cite{schwartz2007space}).

{\bf Acknowledgement: } work by JOR supported by NWO Grant 612.066.826, SMB partly by  NWO Grant 639.021.203.

\small
\bibliographystyle{unsrt}
\bibliography{vidi}

\begin{thebibliography}{10}

\bibitem{fairhall2001multiple}
A.L. Fairhall, G.D. Lewen, W.~Bialek, and R.R.R. van Steveninck.
\newblock {Multiple timescales of adaptation in a neural code}.
\newblock In {\em {NIPS}}, volume~13. The MIT Press, 2001.

\bibitem{wark2009timescales}
B.~Wark, A.~Fairhall, and F.~Rieke.
\newblock {Timescales of inference in visual adaptation}.
\newblock {\em Neuron}, 61(5):750--761, 2009.

\bibitem{panzeri2010sensory}
S.~Panzeri, N.~Brunel, N.K. Logothetis, and C.~Kayser.
\newblock {Sensory neural codes using multiplexed temporal scales}.
\newblock {\em Trends in Neurosciences}, page in press, 2010.

\bibitem{lundstrom2010multiple}
B.N. Lundstrom, A.L. Fairhall, and M.~Maravall.
\newblock {Multiple Timescale Encoding of Slowly Varying Whisker Stimulus
  Envelope in Cortical and Thalamic Neurons In Vivo}.
\newblock {\em J. of Neurosci}, 30(14):50--71, 2010.

\bibitem{vanHateren1997processing}
JH~Van~Hateren.
\newblock {Processing of natural time series of intensities by the visual
  system of the blowfly}.
\newblock {\em Vision Research}, 37(23):3407--3416, 1997.

\bibitem{brenner2000adaptive}
N.~Brenner, W.~Bialek, and R.~de~Ruyter~van Steveninck.
\newblock {Adaptive rescaling maximizes information transmission}.
\newblock {\em Neuron}, 26(3):695--702, 2000.

\bibitem{wark2007sensory}
B.~Wark, B.N. Lundstrom, and A.~Fairhall.
\newblock {Sensory adaptation}.
\newblock {\em Current opinion in neurobiology}, 17(4):423--429, 2007.

\bibitem{famulare2009feature}
M.~Famulare and A.L. Fairhall.
\newblock {Feature selection in simple neurons: how coding depends on spiking
  dynamics}.
\newblock {\em Neural Computation}, 22:1--18, 2009.

\bibitem{drew2006models}
P.J. Drew and LF~Abbott.
\newblock {Models and properties of power-law adaptation in neural systems}.
\newblock {\em Journal of neurophysiology}, 96(2):826, 2006.

\bibitem{lundstrom2008fractional}
B.N. Lundstrom, M.H. Higgs, W.J. Spain, and A.L. Fairhall.
\newblock {Fractional differentiation by neocortical pyramidal neurons}.
\newblock {\em Nature neuroscience}, 11(11):1335--1342, 2008.

\bibitem{hosoya2005dynamic}
T.~Hosoya, S.A. Baccus, and M.~Meister.
\newblock {Dynamic predictive coding by the retina}.
\newblock {\em Nature}, 436:71--77, 2005.

\bibitem{wornell1999signal}
G.W. Wornell.
\newblock {\em {Signal processing with fractals: a wavelet based approach}}.
\newblock Prentice Hall, NJ, 1999.

\bibitem{fusi2005cascade}
S.~Fusi, PJ~Drew, and LF~Abbott.
\newblock {Cascade models of synaptically stored models}.
\newblock {\em Neuron}, 45:1--14, 2005.

\bibitem{bishop1995neural}
C.M. Bishop.
\newblock {\em {Neural networks for pattern recognition}}.
\newblock Oxford University Press, USA, 1995.

\bibitem{chichilnisky2001simple}
EJ~Chichilnisky.
\newblock {A simple white noise analysis of neuronal light responses}.
\newblock {\em Network: Computation in Neural Systems}, 12(2):199--213, 2001.

\bibitem{rieke1999spikes}
F.~Rieke, D.~Warland, and W.~Bialek.
\newblock {\em {Spikes: exploring the neural code}}.
\newblock The MIT Press, 1999.

\bibitem{fairhall2001efficiency}
A.L. Fairhall, G.D. Lewen, W.~Bialek, and R.R.R. van Steveninck.
\newblock {Efficiency and ambiguity in an adaptive neural code}.
\newblock {\em Nature}, 412(6849):787--792, 2001.

\bibitem{kolmogorov1940kurven}
A.~Kolmogorov.
\newblock {Wienersche Spiralen und einige andere interessante kurven in
  Hilbertschen raum}.
\newblock {\em Computes Rendus (Doklady) Academic Sciences USSR (NS)},
  26:115--118, 1940.

\bibitem{mandelbrot1968fractional}
B.B. Mandelbrot and J.W. Van~Ness.
\newblock {Fractional Brownian motions, fractional noises and applications}.
\newblock {\em SIAM review}, 10(4):422--437, 1968.

\bibitem{lundstrom2009sensitivity}
B.N. Lundstrom, M.~Famulare, L.B. Sorensen, W.J. Spain, and A.L. Fairhall.
\newblock {Sensitivity of firing rate to input fluctuations depends on time
  scale separation between fast and slow variables in single neurons}.
\newblock {\em Journal of Computational Neuroscience}, 27(2):277--290, 2009.

\bibitem{pozzorini2010}
C~Pozzorini, R~Naud, S~Mensi, and W~Gerstner.
\newblock Multiple timescales of adaptation in single neuron models.
\newblock In {\em Front. Comput. Neurosci.: Bernstein Conference on
  Computational Neuroscience}, 2010.

\bibitem{xu1996logarithmic}
Z.~Xu, JR~Payne, and ME~Nelson.
\newblock {Logarithmic time course of sensory adaptation in electrosensory
  afferent nerve fibers in a weakly electric fish}.
\newblock {\em Journal of neurophysiology}, 76(3):2020, 1996.

\bibitem{stocker2009}
A~A Stocker and E~P Simoncelli.
\newblock {Visual motion aftereffects arise from a cascade of two isomorphic
  adaptation mechanisms}.
\newblock {\em J. Vision}, 9(9):1--14, 2009.

\bibitem{hong2008intrinsic}
S.~Hong, B.N. Lundstrom, and A.L. Fairhall.
\newblock {Intrinsic gain modulation and adaptive neural coding}.
\newblock {\em PLoS Computational Biology}, 4(7), 2008.

\bibitem{jolivet2006integrate}
R.~Jolivet, A.~Rauch, HR~Luescher, and W.~Gerstner.
\newblock {Integrate-and-Fire models with adaptation are good enough:
  predicting spike times under random current injection}.
\newblock {\em NIPS}, 18:595--602, 2006.

\bibitem{smith2005efficient}
E.~Smith and M.S. Lewicki.
\newblock {Efficient coding of time-relative structure using spikes}.
\newblock {\em Neural Computation}, 17(1):19--45, 2005.

\bibitem{tishby2000information}
N.~Tishby, F.C. Pereira, and W.~Bialek.
\newblock {The information bottleneck method}.
\newblock {\em Arxiv physics/0004057}, 2000.

\bibitem{huys2007fast}
Q.J.M. Huys, R.S. Zemel, R.~Natarajan, and P.~Dayan.
\newblock {Fast population coding}.
\newblock {\em Neural Computation}, 19(2):404--441, 2007.

\bibitem{schwartz2007space}
O.~Schwartz, A.~Hsu, and P.~Dayan.
\newblock {Space and time in visual context}.
\newblock {\em Nature Rev. Neurosci.}, 8(11), 2007.

\end{thebibliography}
\end{document}